\documentclass[a4paper,fleqn]{cas-dc}
\usepackage{graphicx,subcaption}
\usepackage{multicol}
\usepackage{layout}
\usepackage{amsfonts}
\usepackage{amsmath}
\usepackage{amssymb}
\usepackage{textcomp}
\usepackage{caption}
\usepackage[normalem]{ulem}
\usepackage[utf8]{inputenc}
\usepackage{natbib} 
\usepackage{rotating}
\usepackage{lscape}
\usepackage[justification=centering]{caption}
\usepackage{aas_macros}
\usepackage{threeparttable}
\usepackage{epstopdf}
\usepackage{comment}
\usepackage{soul}
\begin{document}
\newcommand{\jetset}{\texttt{JetSet}\xspace}
\newcommand{\NuSTAR}{\textit{NuSTAR}\xspace}
\UseRawInputEncoding
\let\WriteBookmarks\relax
\def\floatpagepagefraction{1}
\def\textpagefraction{.001}
\newcommand{\cdr}[1]{{\bf{\color{red} #1}}}
\newcommand{\cdrtwo}[2]{{\st{#1}}  {{\bf \color{blue} #2}}} 
\shorttitle{3C 454.3}
\shortauthors{Sajeesh.G et al.}

\title[mode = title]{The Ordinary Days of Extraordinary Blazar 3C 454.3}

\author[1,2]{{Sajeesh George}}[orcid=0009-0007-6804-0526]

% Corresponding author indication
\cormark[1]

% Footnote of the first author
%\fnmark[<footnote mark no>]

% Email ID of the first author
\ead{sajichelapurath@gmail.com}

% URL of the first author
%\ead[url]{<URL>}

% Credit authorship
% eg: \credit{Conceptualization of this study, Methodology, Software}
%\credit{<Credit authorship details>}

% Address/affiliation
\affiliation[1]{organization={Govt. College Kodanchery},
%            addressline={Farook College}, 
            city={Calicut},
%          citysep={}, % Uncomment if no comma needed between city and postcode
            postcode={673580}, 
            state={Kerala},
            country={India}} 
\affiliation[2]{organization={University of Calicut},
%            addressline={Thenjipalam}, 
            city={Malappuram},
%          citysep={}, % Uncomment if no comma needed between city and postcode
            postcode={673635}, 
            state={Kerala},
            country={India}}

\author[2,3]{P N {Naseef Mohammed}}

% Corresponding author indication
\cormark[1]

% Footnote of the first author
%\fnmark[<footnote mark no>]

% Email ID of the first author
\ead{naseefmhd@farookcollege.ac.in}

% URL of the first author
%\ead[url]{<URL>}

% Credit authorship
% eg: \credit{Conceptualization of this study, Methodology, Software}
%\credit{<Credit authorship details>}

% Address/affiliation
\affiliation[3]{organization={Farook College},
%            addressline={Farook College}, 
            city={Calicut},
%          citysep={}, % Uncomment if no comma needed between city and postcode
            postcode={673632}, 
            state={Kerala},
            country={India}} 
\author[4]{Aminabi Thekkoth}[]
\ead{athekkoth@ncra.tifr.res.in}
\affiliation[4]{organization={National Centre for Radio Astrophysics (TIFR)},
            %addressline={}, 
            city={Pune},
%          citysep={}, % Uncomment if no comma needed between city and postcode
            postcode={411007}, 
            state={Maharashtra},
            country={India}}  

\author[5]{Vaidehi S. Paliya}[]
\ead{vaidehi.s.paliya@gmail.com}

\affiliation[5]{organization={Inter-University Centre for Astronomy and Astrophysics (IUCAA), SPPU Campus},
            %addressline={}, 
            city={Pune},
%          citysep={}, % Uncomment if no comma needed between city and postcode
            postcode={411007}, 
            state={Maharashtra},
            country={India}}   
\author[2]{C D Ravikumar}[]
\ead{cdr@uoc.ac.in}
         
% Corresponding author text
\cortext[cor1]{Corresponding author}

% Footnote text
%\fntext[1]{}

% For a title note without a number/mark
\nonumnote{}

\begin{abstract}
The blazar 3C~454.3 (\(z = 0.859\)) is one of the brightest and most variable flat-spectrum radio quasars in the sky. Here, we present a multi-wavelength analysis of this enigmatic object in its quiescent states during June 2022 and May 2023 coincident with its first X-ray polarimetric and hard X-ray observations with International X-ray Polarimetric Explorer (\textit{IXPE}) and Nuclear Spectroscopic Telescope Array (\NuSTAR). Though no X-ray polarization was detected, \NuSTAR detected the source on all occasions during this period. We generated five broadband spectral energy distributions (SEDs) of the blazar using quasi-simultaneous observations taken with the \textit{Fermi}-Large Area Telescope, \NuSTAR, and Neil Gehrels \textit{Swift} Observatory. The hard X-ray flux measured by \NuSTAR exhibited mild flux variations. The SEDs were modeled with a one-zone leptonic emission framework using the \jetset package. The optical-ultraviolet spectrum of 3C 454.3 revealed an upturn which we reproduced with the standard accretion disk model. The X-ray to $\gamma$-ray spectrum of 3C~454 was found to be well explained by the external Compton (EC) process with the seed photons provided by the BLR and accretion disk. These results provide intriguing clues about the multi-wavelength behavior of 3C~454.3 during its quiescent activity state.

\end{abstract}


\maketitle
\section{Introduction}

Blazars are a highly energetic subclass of Active Galactic Nuclei (AGN) powered by a central supermassive black hole. They feature powerful relativistic particle jets aimed almost directly along the observer's line of sight on Earth \citep{Blandford1978,1995PASP..107..803U}. Because of this unique orientation, their observed light is dramatically amplified by relativistic beaming, making them the most luminous persistent objects in the universe. They are distinguished by rapid variability and high polarization across the entire electromagnetic spectrum, from radio waves to gamma rays \citep{1995PASP..107..803U}. Blazars are spectroscopically classified into two main subclasses based on the properties of their optical emission lines: Flat Spectrum Radio Quasars (FSRQs) and BL Lacertae (BL Lac) objects. FSRQs are observationally characterized by strong broad lines with a rest-frame equivalent width, \mbox{\(\text{EW} \ge 5\text{ \AA}\)}. Physically, this corresponds to a  broad-line region (BLR) luminosity of \(L_{\text{BLR}}/L_{\text{Edd}}  > 5 \times 10^{-4}\) \citep{Ghisellini2011}, where \(L_{\text{Edd}}\) is the Eddington luminosity. BL Lacs show weak or absent lines ($\text{EW} < 5$\,\AA) due to radiatively inefficient accretion and a jet continuum that dilutes thermal features \citep{Bottcher2019}. Despite these central engine differences, both classes display shared jet behaviors, including rapid flux and polarization variability \citep{Wagner1995, Andruchow2005}, flat radio spectra and superluminal patterns \citep{Jorstad2005}.

The origin of the high-energy component in the characteristic two-hump 
spectral energy distribution (SED) of blazars remains a subject of intense debate. In conventional leptonic frameworks, this emission feature is attributed to inverse Compton (IC) scattering by relativistic electrons. This occurs either via synchrotron self-Compton (SSC) processes involving internal synchrotron photons \citep{konigl1981, ghisellini1985, maraschi1992, bloom1996}, or through external  Compton (EC) scattering \citep{sikora1994, blazejowski2000, ghisellini2009}. The external seed photon fields vary with distance from the central engine and may include thermal contributions from the accretion disc \citep{dermer1992, dermer1993}, the BLR \citep{sikora1994}, or the dusty torus \citep{blazejowski2000}. In the context of hadronic scenarios, the high-energy spectral component is attributed to protons accelerated to ultra-relativistic velocities within the jet. This emission feature is modeled either as direct proton synchrotron radiation or as radiative output from secondary leptons-such as muons and electron-positron pairs-produced via photo-hadronic (\(p\gamma\)) interactions between the energetic protons and ambient low-energy photon fields \citep{mannheim1993, dermer2001, muecke2001}. Based on the classification scheme proposed by \citet{Abdo_2010}, blazars are divided into distinct categories using their synchrotron peak frequency ($\nu_{\text{peak}}^S$). These sources are designated as Low-Synchrotron-Peaked (LSP) when $\nu_{\text{peak}}^S < 10^{14}$~Hz, Intermediate-Synchrotron-Peaked (ISP) when $10^{14}\text{ Hz} \le \nu_{\text{peak}}^S \le 10^{15}$~Hz, and High-Synchrotron-Peaked (HSP) when $\nu_{\text{peak}}^S > 10^{15}$~Hz.

The object 3C~454.3, located at a redshift of \(z = 0.859\) \citep{jackson1991optical}, stands out as a highly luminous and dynamic FSRQ, renowned for its intense brightness and extreme fluctuations in energy output. In April--May 2005, the source underwent a monumental optical outburst, reaching a historic peak magnitude of \(R = 12.0\), making it the most luminous quasar state ever observed at the time \citep{villata2006unprecedented}. Following its launch, the \textit{AGILE} satellite frequently detected 3C~454.3 as the brightest \(\gamma\)-ray blazar in the sky. Furthermore, data from the \textit{Fermi} Gamma-ray Space Telescope recorded a series of historic \(\gamma\)-ray flares in late 2009, during which the object temporarily became the brightest persistent \(\gamma\)-ray source in the sky \citep{vercellone2011gamma}, surpassing even the Vela Pulsar. During its December 2009 and November 2010 outbursts, its daily integrated \(\gamma\)-ray flux (\(E > 100\,\text{MeV}\)) surpassed the extraordinary threshold of \(\sim 10^{-5}\,\text{photons\, cm}^{-2}\,\text{s}^{-1}\), peaking at \((2.0 \pm 0.4) \times 10^{-5}\,\text{photons\, cm}^{-2}\,\text{s}^{-1}\) and \((6.8 \pm 1.0) \times 10^{-5}\,\text{photons\, cm}^{-2}\,\text{s}^{-1}\), respectively \citep{Striani2010, Vercellone2011}. The extreme brightness amplitudes of 3C~454.3 have been a major focus of multi-wavelength tracking campaigns. During its historically high active phases, particularly during massive \(\gamma\)-ray flaring events, synchronized observations across optical, X-ray, and GeV bands have confirmed tight flux correlations. Modeling of these peak periods indicates that the intense \(\gamma\)-ray emission varies quadratically or higher with the optical flux. This behavior can be successfully explained via a simple one-zone synchrotron inverse Compton (IC) framework where Doppler boosting factors increase significantly in a fast-moving jet region \citep{bonnoli2011gamma, sahakyan2021modelling}. This interpretation is further corroborated by observations where the \(\gamma\)-ray flux increases by a factor of 10 against mild baseline variations, pointing towards an external Compton (EC) component driven by a local enhancement of soft seed photons \citep{2011ApJ...736L..38V}. Investigating these low-energy baseline dynamics, systematic studies targeting the quiescent states of bright FSRQs revealed that the high-energy index of the electron distribution scales directly with the observed \(\gamma\)-ray luminosity. This demonstrates that the kinetic power carried by electrons remains sufficient to sustain jet radiation fields even when the blazar is in a non-flaring state \citep{2021MNRAS.504.1103R}. The spectral break observed in the MeV--GeV \(\gamma\)-ray regime of 3C~454.3 during extreme flaring epochs remains a subject of ongoing debate between extrinsic absorption and intrinsic particle distribution models. While early analyses by \citet{Cerruti2013} attributed this feature to Klein--Nishina scattering effects under equipartition, recent multi-wavelength modeling by \citet{Kang2021} points toward an inherent broken power-law electron distribution within a two-component EC framework. Broadband spectral energy distribution (SED) modeling has further refined these constraints; while localized synchrotron self-Compton (SSC) and EC mechanisms successfully account for isolated high-activity windows like the intense August 2015 flaring state \citep{Shah2017}, long-term multi-wavelength correlation and structural variability studies indicate that systemic modifications to the Doppler factor, electron distribution hardening, and localized magnetic reconnection events dominate its broad emission profile \citep{Zhou2021, Paliya2021}.

The Nuclear Spectroscopic Telescope Array (\NuSTAR) observed the bright blazar 3C 454.3 in June 2022 and May 2023. These observations were part of a larger project with the Imaging X-ray Polarimetry Explorer (\textit{IXPE}) to study how the blazar's light is polarized. The source was in a low-activity state, and no polarization was detected \citep[see,][for \textit{IXPE} results]{2024ApJ...972...74M}. However, these data represent the first \NuSTAR observations of 3C 454.3. We study its multi-band properties here, focusing on the \NuSTAR data, including its variability and SED modeling. The structure of this paper is organized as follows. Section 2 outlines the observational data collected from the \textit{Fermi}, \textit{Swift}, and \textit{NuSTAR} space telescopes, along with the corresponding data reduction techniques. In Section 3, we describe the SED modeling Python package, the \jetset framework. A detailed analysis 
and interpretation of our findings are presented in Section 4 and Section 5. 
Section 6 provides a summary of the study. 
Throughout this work, we assume a flat cosmological model parameterized by 
$\Omega_m = 0.3$ and $H_0 = 70$~km~s$^{-1}$~Mpc$^{-1}$.

\section{Data Reduction}\label{sec:data_red}
\subsection{Fermi-LAT}

The \textit{Fermi-Large Area Telescope} (LAT) is a high-energy \(\gamma\)-ray detection telescope operating in the energy range 20 MeV to over 300 GeV. It covers 20 percent of the sky at any time \citep{2009ApJ...697.1071A}. In this work, we collected data from the NASA database for the two periods: Period I: MJD 59740-59747 (2022 June 10-17) and Period II: MJD 60064-60080 (2023 April 30 - May 16). The LAT data analysis was done using \textit{fermipy} version 1.4.0 \citep{wood2017fermipyopensourcepythonpackage} as per the standard procedure mentioned in the \textit{Fermi}-LAT documentation. A circular region with a $15^{\circ }$ radius was used for source events extraction, and a zenith angle cut of less than $90^{\circ }$ was applied to exclude data from the Earth's limb. A $10^{\circ }$ radius was selected as the region of interest (ROI), and the parameters of sources outside this ROI were frozen to their respective catalog values during the analysis. The diffuse background model was created using the {gll\_iem\_v07} file along with the {iso\_P8R3 \_SOURCE \_V3 \_v1} files. The XML model file for the source was created using the 4FGL catalog \citep{2020ApJS..247...33A}. The source 3C 454.3 was modeled by a simple power-law (PL)  model to generate light curves and SEDs. The PL indices obtained from this model show smaller statistical uncertainties when compared to those obtained from more complex models. Furthermore, adopting a simple PL model is appropriate since our goal is to probe the shortest timescales, which inherently limits photon statistics. To evaluate the statistical significance of the \(\gamma \)-ray signal, we utilized the maximum likelihood (ML) test statistic, defined as \(\text{TS} = 2\Delta\ln(\mathcal{L})\), where \(\mathcal{L}\) represents the likelihood ratio between models with and without a target \(\gamma \)-ray point source. For the spectral analyses, we consider a source to be detected if its Test Statistic (TS) exceeds 9, which corresponds to a \(\sim 3\sigma\) detection significance \citep{1996ApJ...461..396M}.

\subsection{NuSTAR}

\NuSTAR is a space telescope capable of detecting high-energy X-rays within the energy range of 3-79 keV. The blazar 3C454.3 was targeted by \NuSTAR once in June 2022 and on four separate occasions in May 2023, accumulating exposures above 40 ks per epoch.\footnote{Due to low Earth orbit, the net exposure time is roughly 50\% of the total observation length.} We collected the data for the above five observations from the \NuSTAR data archive.\footnote{\url{https://heasarc.gsfc.nasa.gov/docs/nustar/nustar_archive.html}} We used the \NuSTAR Data Analysis Software version 1.10 and standard pipeline for the data analysis.\footnote{\url{https://heasarc.gsfc.nasa.gov/docs/nustar/analysis/nustar_swguide.pdf}} We used the \texttt{nuproducts} tool to extract light curves and spectra for both focal plane modules (FPMA and FPMB). For the source spectra, we used a $30''$ circular extraction region centered on the source 3C454.3. To extract the background spectra, we selected a source-free circular region with a $70''$ radius on the same detector chip. Finally, we grouped the source spectra to get a valid spectral fit. In addition, the final light curves were produced by merging the count rates from both FPMA and FPMB detectors. This process included background subtraction and the application of a $1\text{~ks}$ time binning.

\subsection{Swift}

The \textit{Swift} satellite observed the blazar 3C 454.3 seven times during the Period I, and five times during the Period II. We utilized the automated online \textit{Swift}-XRT data product generator \citep{2009MNRAS.397.1177E} to analyze the XRT data and to create the observation-binned light curves. For SED modeling, the source and background spectra for the selected epochs were generated using the same online pipeline. Spectral analysis was performed by loading the extracted data into XSPEC \citep{1996ASPC..101...17A}, where the spectra were fitted with a power-law model to derive the X-ray fluxes. To correct for Galactic absorption, the neutral hydrogen column density (\(N_{\text{H}}\)) was fixed at \(8.59 \times 10^{20}\text{ cm}^{-2}\), adopted from the LAB survey \citep{refId0}.

We retrieved the \textit{Swift}-UVOT data from the HEASARC archive and processed them following standard data reduction protocols. For each available band, individual image frames were co-added using the HEASOFT task \texttt{uvotimsum}. Photometric data extraction was conducted using \texttt{uvotsource}, adopting a \(5^{\prime \prime }\) circular aperture centered on the target, while a nearby, source-free \(20^{\prime \prime }\) circular region provided the background baseline. To account for Galactic interstellar dust attenuation, we applied an extinction correction based on a reddening value of \(E(B-V) = 0.0932\) and a standard total-to-selective extinction ratio of \(R_V = 3.1\) \citep{Schlafly_2011}.
\subsection{Archival Observations}

We collected the archival data from the SSDC SED builder http://tools.asdc.asi.it/SED/\citep{2011arXiv1103.0749S}.

\section{SED modelling using JetSet  }\label{sec:model}

To simulate the broadband multi-epoch spectrum, we utilize \jetset, an open-source C/Python framework designed to model radiative and accelerative processes in relativistic jets \citep{Tramacere_2011,Tramacere}. The platform numerically handles diverse observational datasets via \texttt{astropy} bindings to resolve complex leptonic scenarios, including synchrotron self-Compton (SSC) and external Compton (EC) scattering configurations.

The model assumes the emission region to be a spherical blob of plasma with radius \(R\) at a distance (\(R_{\rm H}\)) from the central supermassive black hole of mass $M_{BH}$. This region travels along the jet at relativistic speed with a bulk Lorentz factor \( \Gamma \), at a very small viewing angle of (\(\theta \)) with respect to the line of sight of the observer.  The observed flux at Earth will be Doppler-boosted by the kinematic Doppler factor (\(\delta \)) \citep{1995PASP..107..803U}. The magnetic field $B$ associated with the emission region is assumed to be uniform and tangled.
%The physical parameters applied during the SED modeling process are summarized in Table 6.
 The model utilizes the accelerated electron population characterized by a broken power-law distribution given by:
\begin{equation}
n(\gamma) = 
\begin{cases}
k \gamma^{-p}, & \gamma_{\rm min} \le \gamma \le \gamma_{\rm break} \\ 
k \gamma_{\rm break}^{(p_{1}-p)} \gamma^{-p_{1}}, &  \gamma_{\rm break} \le \gamma \le \gamma_{\rm max}
\end{cases}
\end{equation}
 Where $\gamma$ is the dimensionless electron energy and $\gamma_{\rm min}$ and $\gamma_{\rm max}$ are its minimum and maximum values. \(p\) and \(p_{1}\) represent the spectral indices below and above the break Lorentz factor \(\gamma _{\text{break}}\), respectively. The \(\gamma _{\text{break}}\) corresponds to the energy of electrons whose radiative-cooling timescale is equal to the effective residence time of the electrons in the emission region. These particles radiate through synchrotron and inverse Compton processes in the magnetic field.

The physical geometry of the emission region is constrained following the structural baselines established for \mbox{3C 454.3} by \cite{bonnoli2011gamma} and \cite{2021MNRAS.504.1103R}. The radius of the spherical plasma blob is fixed at \(R = 1.3 \times 10^{16}\) cm. Assuming a strict conical jet profile where the emission region completely fills the jet cross-section, the distance of the dissipation zone from the central supermassive black hole is locked at \(R_H = 1.6 \times 10^{17}\) cm. To ensure physical consistency, this radius was back-evaluated against the causality constraint \(t_{\text{var}} \ge 2R(1+z)/(c\delta)\); given a calculated Doppler factor of \(\delta \approx 31\)-\(33\), this zone yields a minimum variability limit of \(\sim 14\) hours. 
%This easily accommodates the multiwavelength flux variations tracked across our selected integration epochs, which span 2 to 6 days in length. 
The jet viewing angle (\(\theta \)) is fixed at  \(1.3^{\circ }\) \citep{2009A&A...498..723H}. The accretion disk luminosity (\(L_{\text{d}}\)) is taken as \(2.24 \times 10^{46}\text{ erg s}^{-1}\), and the black hole mass is \(8.9 \times 10^8 M_{\odot}\) from optical spectroscopy  \citep{Paliya_2021}. Following the scaling relations proposed by \citet{Ghisellini2010},
%established via reverberation mapping techniques 
we determined the distances of the broad-line region ($R_{\text{BLR}}$) and the dusty torus ($R_{\text{DT}}$) from the central engine using the accretion disk luminosity ($L_{\text{d}}$):
\begin{equation}
R_{\text{BLR}} = 10^{17} \times \left(\frac{L_{\text{d}}}{10^{45}}\right)^{1/2}
\end{equation}
\begin{equation}
R_{\text{DT}} = 2.5 \times 10^{18} \times \left(\frac{L_{\text{d}}}{10^{45}}\right)^{1/2}
\end{equation}
Substituting the value of the disk luminosity into the two equations yields a broad-line region radius $R_{\text{BLR}} = 4.73 \times 10^{17}$~cm and a Dusty Torus radius $R_{\text{DT}} = 11.8 \times 10^{18}$~cm. Treating the broad-line region as a thin, spherical shell composed of ionized gas, its position acts as the midpoint between its inner ($R_{\text{in}}$) and outer ($R_{\text{out}}$) boundaries, such that $R_{\text{BLR}} = (R_{\text{in}} + R_{\text{out}})/2$. The total physical thickness of this shell is fixed at $R_{\text{out}} - R_{\text{in}} = 2 \times 10^{16}$~cm \citep{2009MNRAS.397..985G}. The temperatures of the accretion disk and the dusty torus were fixed at $30,000$~K and $900$~K, respectively. The fraction of disk luminosity reprocessed by the broad-line region ($\tau_{\text{BLR}}$) and the fraction absorbed and re-emitted by the dusty torus ($\tau_{\text{DT}}$) were both set at 0.1. All the remaining parameters were calculated from the SED fitting. 

\section{Results}\label{sec:results}
In Figure 1, we present the multi-wavelength light curves for Period I and Period II. These epochs were selected on the basis of the \NuSTAR detections and the availability of \textit{Swift} XRT/UVOT observations. The one-day binned \(\gamma \)-ray light curve is shown in the top panel, followed by the observation-binned X-ray (0.3-10 keV) light curve in the second panel. The UV and optical light curves are presented in the third and fourth panels, respectively. Figure 2 shows the hard X-ray light curves (3-79 keV) obtained from the five \NuSTAR detections. Significant multi-wavelength flux variations are evident across all observed energy bands. The fractional variability amplitudes (\(F_{\text{var}}\)) for both periods are presented in Table 1.
 
To investigate the underlying radiative mechanisms driving the observed emission patterns, we constructed broadband SEDs for the five distinct epochs. These SED modeling epochs were chosen based on the timing of \NuSTAR detections and the availability of \(\gamma \)-ray flux measurements that yield a TS value greater than 25.  Details of all the selected epochs are given in Table 2. The spectral parameters obtained after generating SED points from \textit{Fermi}-LAT, \textit{Swift} XRT/UVOT and \NuSTAR are given in Table 3. The parameters used and derived from the SED modeling are given in Table 4. 
%The energy densities and powers obtained after the SED modeling are given in Table 5.

\begin{figure*}[t]
\centering

\begin{subfigure}{0.48\textwidth}
    \includegraphics[width=\linewidth]{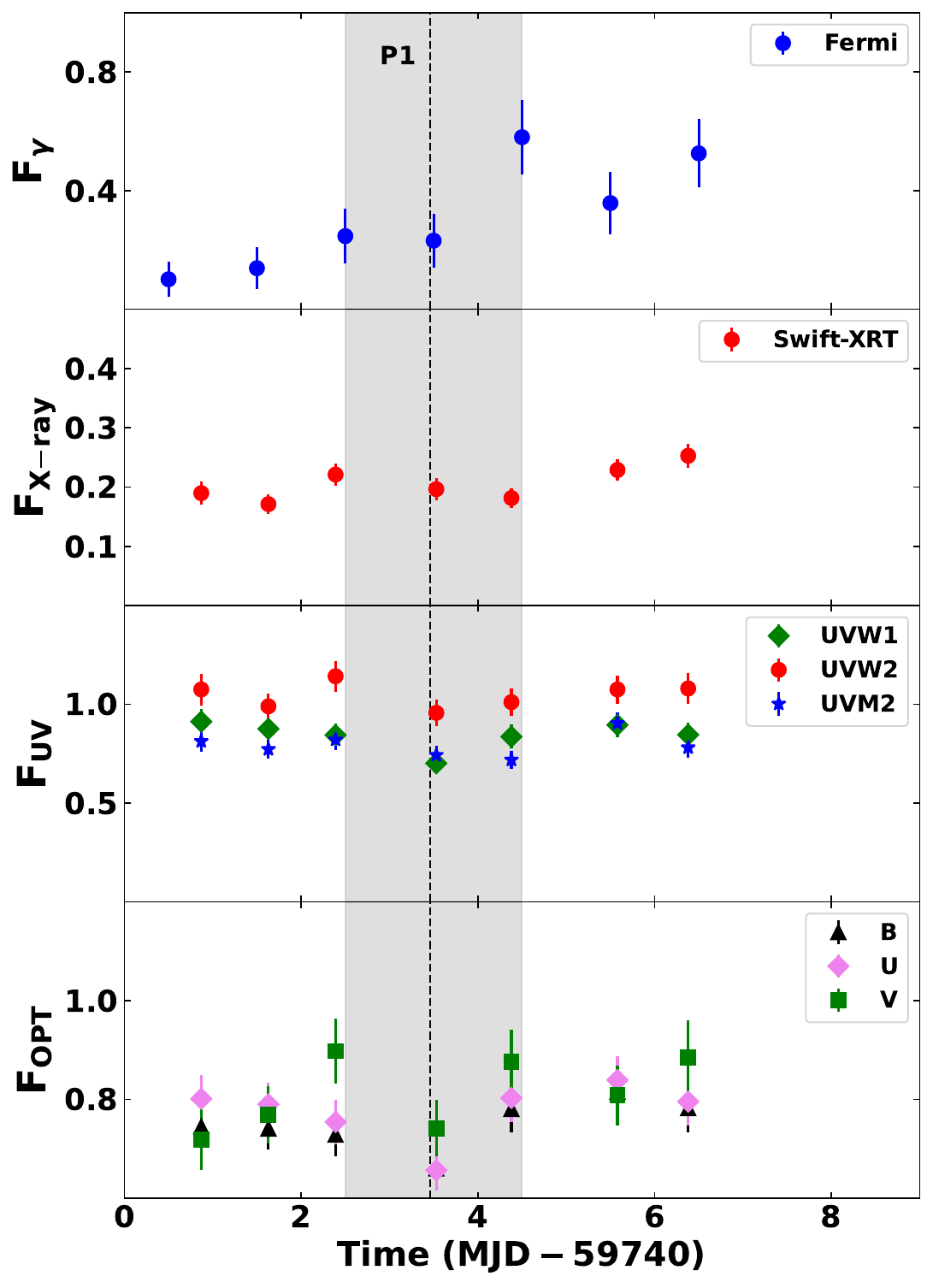}
    \end{subfigure}
\hfill
\begin{subfigure}{0.48\textwidth}
    \includegraphics[width=\linewidth]{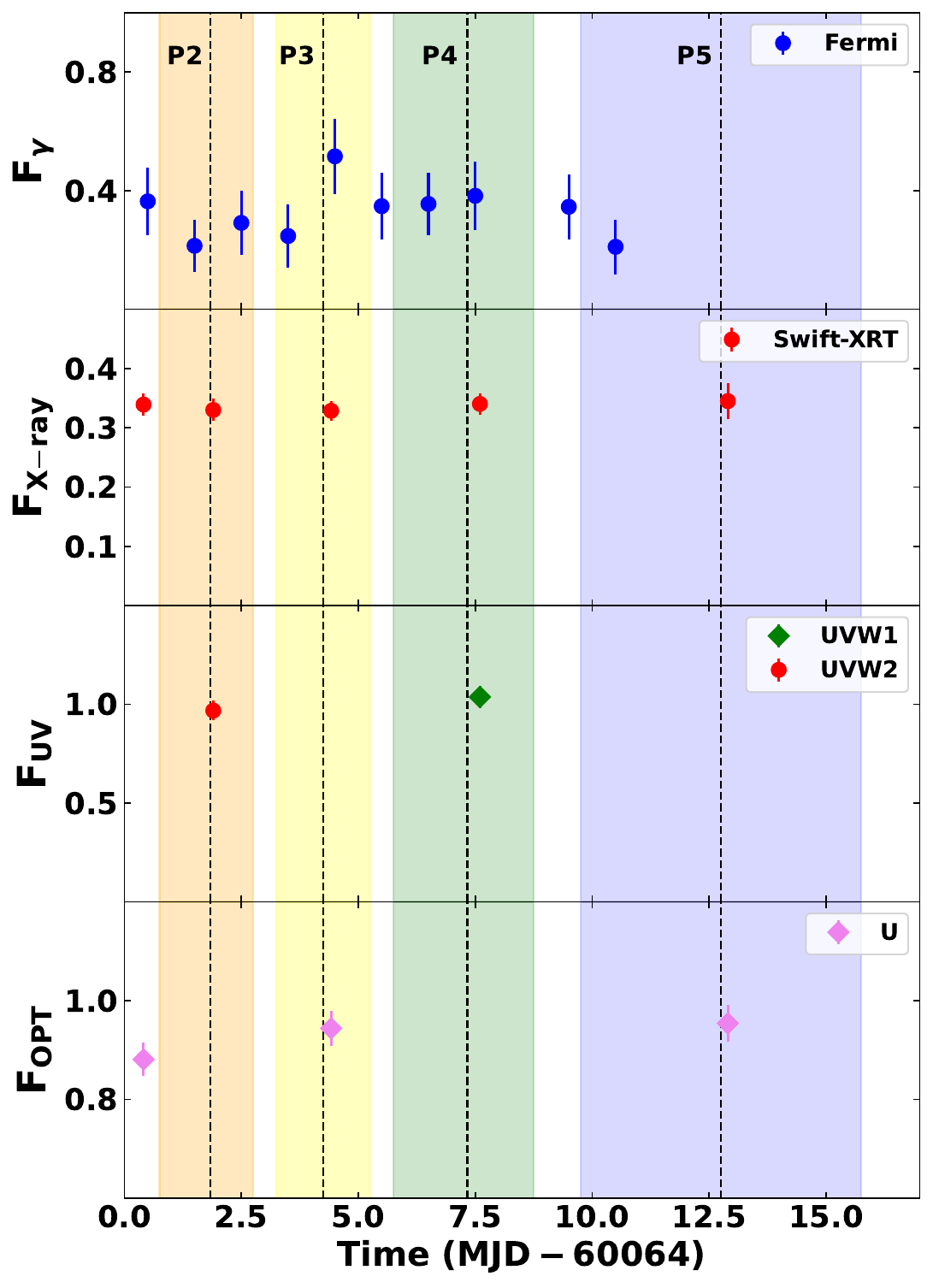}
    \end{subfigure}

\caption{Multi-wavelength light curves of 3C 454.3. One day binned \textit{Fermi}\text{--}Lat data points are in units of $10^{-6}$ ph\ cm$^{-2}$\ s$^{-1}$, \textit{Swift}\text{--}XRT in Count/seconds, \textit{Swift}\text{--}UV and \textit{Swift}\text{--}Optical data points are in units of $10^{-11}$ erg\ cm$^{-2}$\ s$^{-1}$. The dotted lines indicate \textit{NuSTAR} detections. The colored shaded regions highlight the epochs selected for broadband SED generation and modeling.}
\label{multi-wavelength lightcurve}
\end{figure*}

\begin{figure*}[t]
\centering
\hbox{
     \includegraphics[scale=0.35]{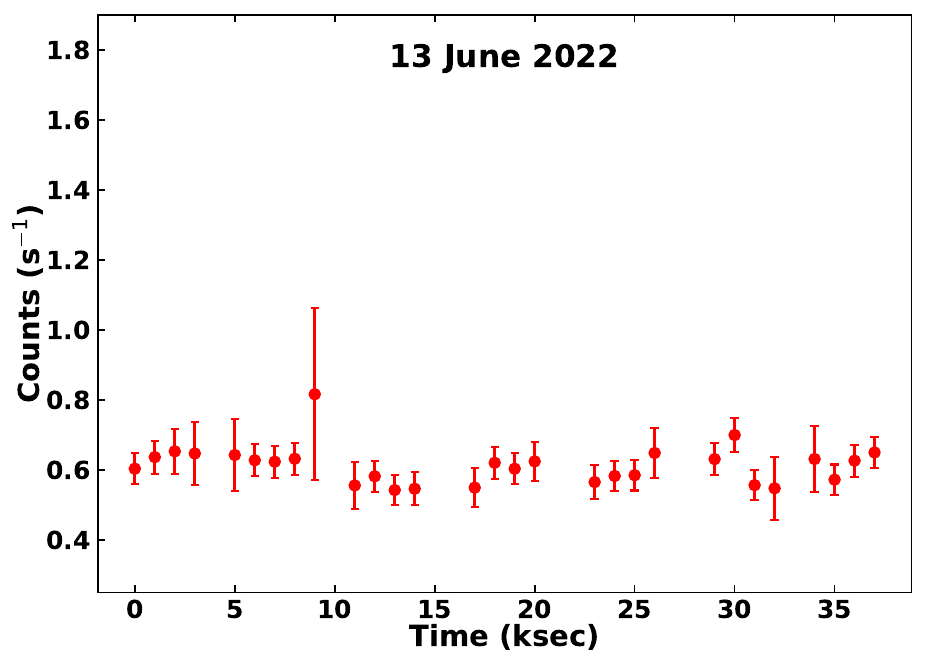}
    \includegraphics[scale=0.35]{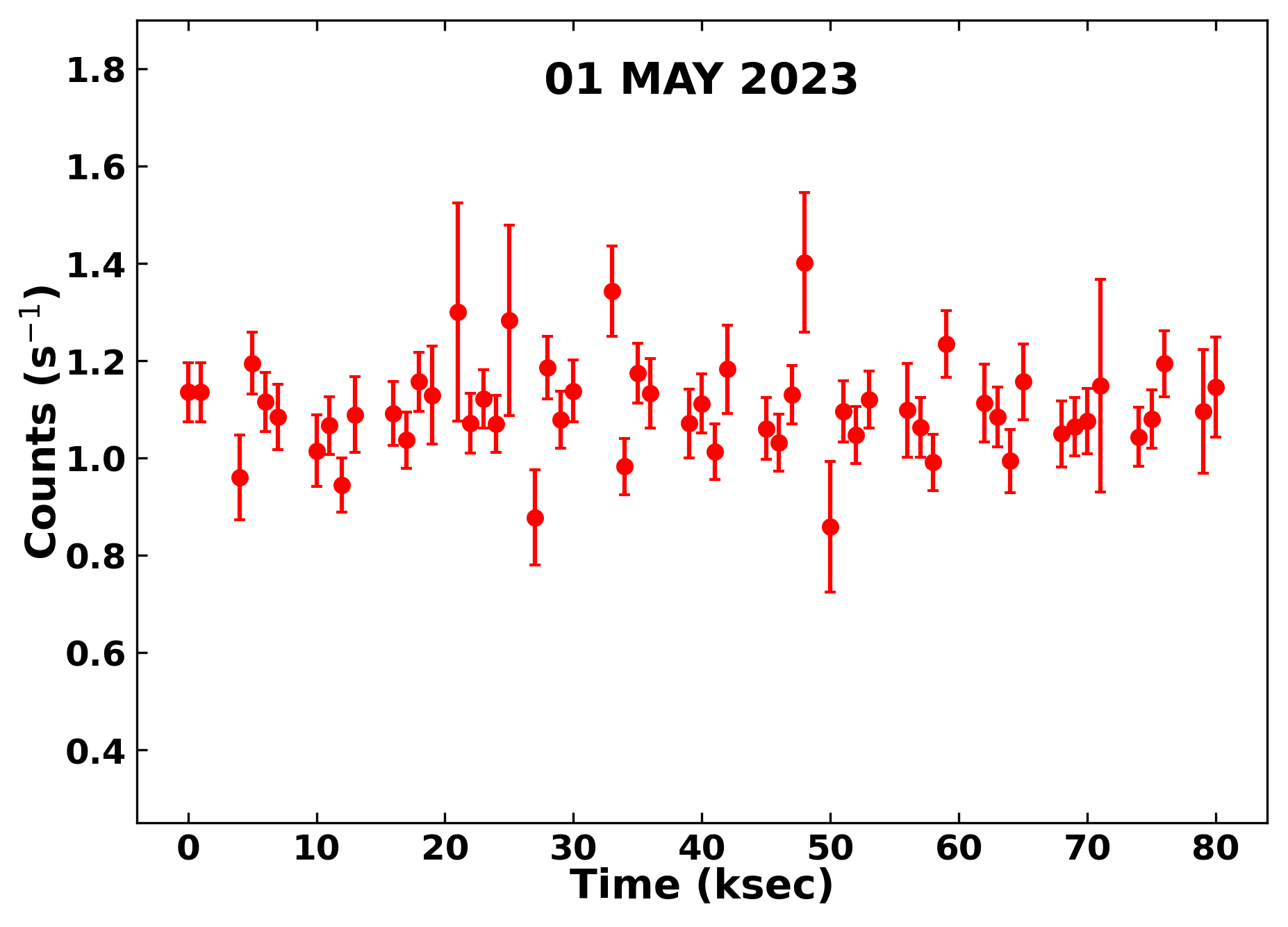}
    \includegraphics[scale=0.35]{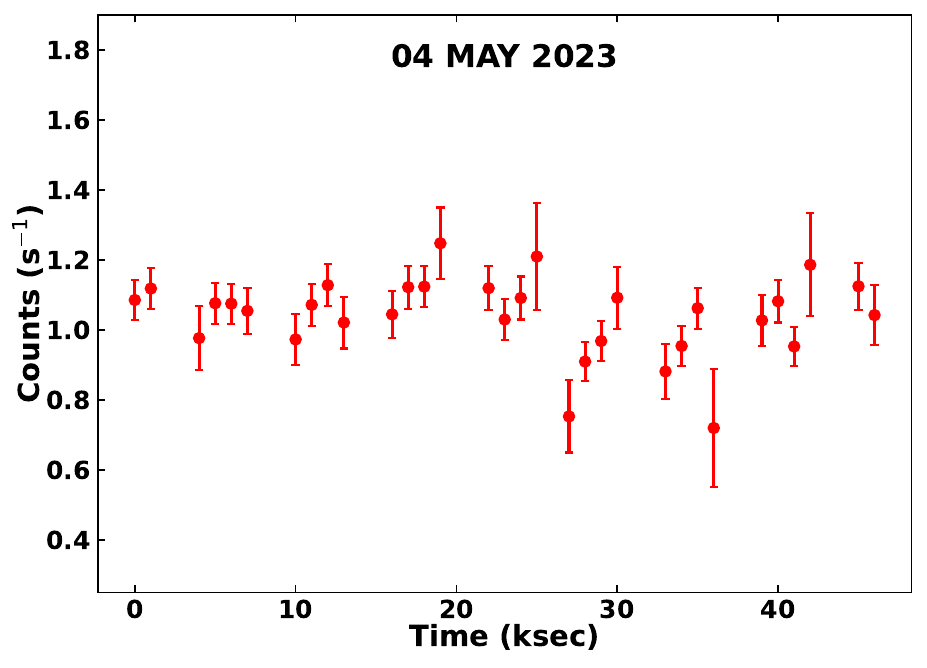}   
}
\hbox{\hspace{2.9cm}
     \includegraphics[scale=0.35]{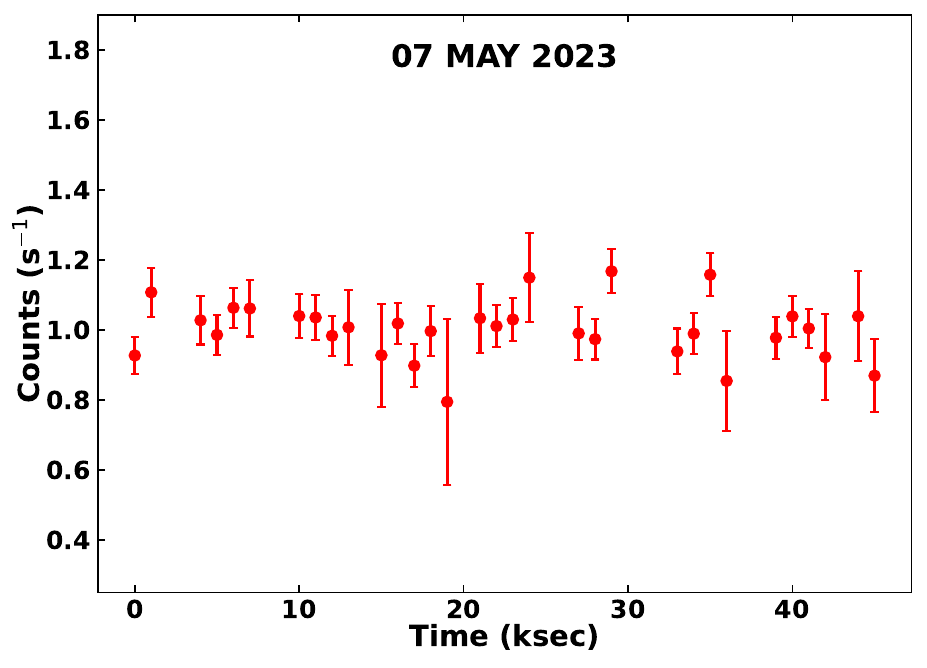}
     \includegraphics[scale=0.35]{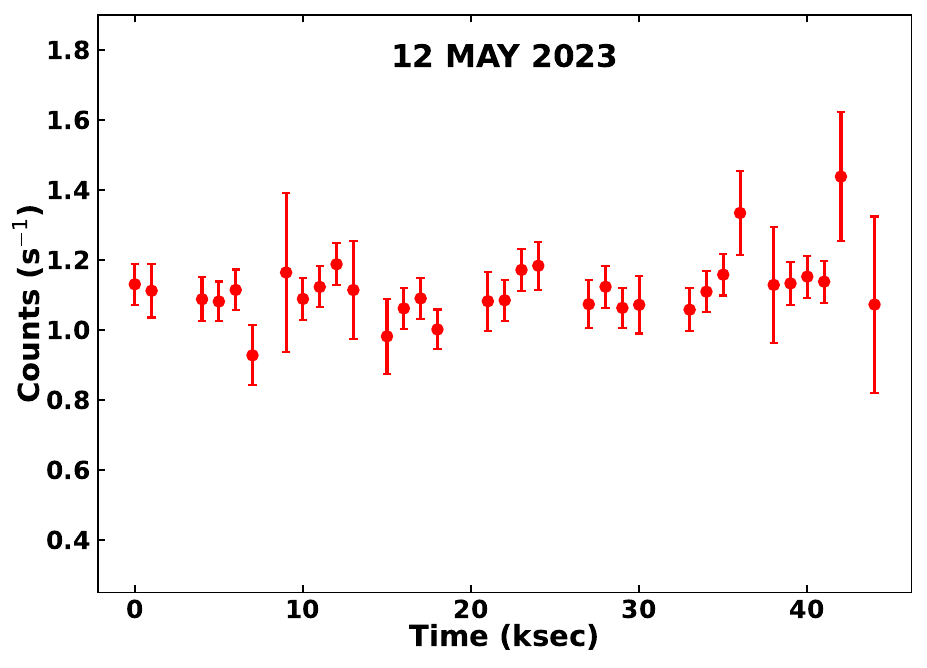}
}
\caption{NuSTAR background-subtracted 3-- 79 keV light curves of the five observations. For this plot, the FPMA and FPMB count rates are
summed, and 1 ks binning is applied.}
\label{NuSTAR_light_curves}
\end{figure*}

\begin{table*}[htbp]
\centering
\caption{Fractional rms Variability Amplitude Parameter ($F_{\text{var}}$), calculated for the Light Curve Shown in Figure 1}
\label{tab:variability}
\vspace{0.5em}
\small
\begin{tabular*}{\textwidth}{@{\extracolsep{\fill}}lcc}
\hline\hline
Energy Band & \makecell[c]{$F_{\text{var}}$ \\ (MJD 59740--59747)} & \makecell[c]{$F_{\text{var}}$ \\ (MJD 60064--60080)} \\ [0.5ex]
\hline
$\gamma$-ray ($0.1$--$300\text{ GeV}$) & $0.51 \pm 0.13$ & $-$ \\
XRT ($3$--$79\text{ keV}$)             & $0.14 \pm 0.02$ & $-$ \\
XRT ($0.3$--$10\text{ keV}$)            & $0.11 \pm 0.04$ & $0.03 \pm 0.04$ \\
UVW2                                    & $0.04 \pm 0.03$ & $-$ \\
UVM2                                    & $-$             & $-$ \\
UVW1                                    & $0.04 \pm 0.04$ & $-$ \\
U                                       & $0.05 \pm 0.03$ & $0.02 \pm 0.04$ \\
B                                       & $0.03 \pm 0.04$ & $-$ \\
V                                       & $0.05 \pm 0.05$ & $-$ \\ [0.5ex] 
\hline % Single line at the bottom
\end{tabular*}
\end{table*}

\begin{table}
\caption{Different epochs selected for the broadband SED generation and modeling.}
\setlength{\tabcolsep}{33pt}
\renewcommand{\arraystretch}{1.2}
\begin{tabular}{llc}
    \hline\hline
\text{Epoch}          & \text{Time(MJD)} \\
    \hline
P1           & 59742.50\text{--}59744.50         \\
P2 & 60064.75\text{--}60066.75       \\
P3   & 60067.25\text{--}60069.25      \\
P4        & 60069.75\text{--}60072.75      \\
P5         & 60073.75\text{--}60079.75             \\
\hline
    \end{tabular}
    \label{epochs}
\end{table}

\begin{table*}
	\centering
	\setlength{\tabcolsep}{30pt}
	%\scriptsize
	\caption{The summary of the data analysis performed for the \textit{Fermi}-LAT, \textit{Swift} XRT/ UVOT and \textit{NuSTAR} observations.Integrated $\gamma$-ray flux is in units of \mbox{$10^{-7}$ ph\ cm$^{-2}$\ s$^{-1}$}. \textit{Swift} XRT/ UVOT and \textit{NuSTAR}-XRT fluxes are in units of $10^{-12}$ ergs cm$^{-2}$ s$^{-1}$.}
	\label{tab:Fermi lat swift uvot-Xrt NuSTAR}

    \begin{tabular*}{\textwidth}{@{\extracolsep{\fill}}lccc}
        \hline\hline
        \multicolumn{4}{c}{\textit{Fermi}-LAT } \\
        \hline
        Epoch &Flux& Photon Index & TS \\
        \hline
        P1& 3.86$\pm$1.09  & 2.31$\pm$0.20           & 56.01      \\
        P2& 4.10$\pm$1.22  & 2.62$\pm$0.27           & 30.49      \\
        P3& 3.83$\pm$1.25  & 2.33$\pm$0.23           & 47.15      \\
        P4& 3.50$\pm$0.93  & 2.40$\pm$0.22           & 54.55      \\
        P5& 2.11$\pm$0.30  & 2.52$\pm$0.08            &  49.39\\    \hline 
	\end{tabular*}
    \setlength{\tabcolsep}{14pt}
	
	\vspace{3pt} % Adds some space between sections
	% Swift-XRT Section (Modified to 5 Columns)
	\begin{tabular*}{\textwidth}{@{\extracolsep{\fill}}lccc}
         
		\multicolumn{4}{c}{\emph{Swift}-XRT } \\
		\hline
		Epoch & Flux & Photon Index ($\Gamma$/$\alpha$) & $\chi^{2}$/Dof         \\       
		\hline
		P1 & $6.46\pm 0.60$ & $1.56 \pm 0.10 $ & 42.23/42 \\
		P2 & $16.5\pm 1.40$ & $1.20 \pm 0.09$ & 24.06/30 \\
		P3 & $14.7\pm 0.97$ & $1.46 \pm 0.08$ & 23.44/24 \\
        P4 & $15.2\pm 1.20$ & $1.30 \pm 0.08$ & 55.12/64 \\ 
        P5  & $14.2\pm 1.10$ & $1.29 \pm 0.14$ & 10.61/11 \\
		\hline
	\end{tabular*}
    \setlength{\tabcolsep}{14pt}
	\vspace{3pt} % Adds some space between sections
	  
	\begin{tabular*}{\textwidth}{@{\extracolsep{\fill}}lcccccc}
		\multicolumn{7}{c}{\emph{Swift}-UVOT (Flux) }\\
		\hline
		Epoch&V &B& U& UVW1& UVM2& UVW2\\
		\hline
		P1 & 8.00 $\pm$ 0.43 & 7.12 $\pm$ 0.31 & 7.17 $\pm$ 0.31 & 7.58 $\pm$ 0.39 & 9.83 $\pm$ 0.48 & 7.31 $\pm$ 0.33\\
		P2 & - & -& - & -& - & 9.45 $\pm$ 0.24 \\ 
		P3 & - & - & 9.29 $\pm$ 0.21 & - & - & - \\
		P4 & - & - & - &  & 10.40 $\pm$ 0.24 & - \\
		  P5 & - & - & 9.41 $\pm$ 0.24 & - & - & - \\ 
	    \hline
	\end{tabular*}
    \setlength{\tabcolsep}{14pt}
    \vspace{3pt}

    \begin{tabular*}{\textwidth}{@{\extracolsep{\fill}}lccc}

		\multicolumn{4}{c}{\emph{NuSTAR}-XRT} \\
		\hline
		Epoch & Flux  & Photon Index ($\Gamma$/$\alpha$) & $\chi^{2}$/Dof         \\       
		\hline
		P1 & $30.15\pm 0.97$ & $1.47 \pm 0.03 $ & 208.21/207 \\
		P2 & $54.67\pm 0.92$ & $1.47 \pm 0.01$ & 549.97/557 \\
		P3 & $52.38\pm 1.20$ & $1.46 \pm 0.02$ & 375.67/334 \\
        P4 & $51.99\pm 1.20$ & $1.44 \pm 0.02$ & 273.00/299 \\ 
        P5  & $52.97\pm 1.14$ & $1.50 \pm 0.02$ & 361.30/362 \\
		\hline
    \end{tabular*}

\end{table*}

\begin{table*}[!htbp]
\centering
\caption{The modeled SED parameters for the five activity periods (P1--P5). See Section~\ref{sec:model} for details.}
\label{tab:jetset_parameters}
\small % Sets a uniform, readable font size
\setlength{\tabcolsep}{0pt} % Let LaTeX handle spacing automatically
\begin{tabular*}{\textwidth}{@{\extracolsep{\fill}}lccccc}
\hline\hline
Parameter & P1 & P2 & P3 & P4 & P5 \\
\midrule
%$R$ ($10^{16}$ cm)                     & 1.3 & 1.3 & 1.3 & 1.3 & 1.3 \\
%Viewing angle, $\theta$ (deg)          & 1.3    & 1.3 & 1.3 & 1.3 & 1.3 \\
$N$                                    & 2200 & 3500 & 3400 & 3300 & 4700 \\
$\gamma_{\rm min}$                     & 1.8    & 1.9 & 1.8 & 1.6 & 1.7 \\
$\gamma_{\rm max}$                     & 10000 & 6000 & 9000 & 7000 & 7500 \\
$\gamma_{\rm break}$                   & 55 & 40 & 21 & 24 & 25 \\
$p$                                    & 1.8 & 1.6 & 1.6 & 1.6 & 1.6 \\
$p_{1}$                                & 3.4 & 3.65 & 3.2 & 3.35 & 3.4 \\
Bulk Lorentz Factor ($\Gamma$)         & 18.2 & 19.2 & 19.0 & 19.9 & 17.5 \\
%$L_{\rm disk}$ ($10^{46}$ erg s$^{-1}$)& 2.24 & 2.24 & 2.24 & 2.24 & 2.24 \\
%$R_{\rm BLR,in}$ ($10^{17}$ cm)        & 4.63 & 4.63 & 4.63 & 4.63 & 4.63 \\
%$R_{\rm BLR,out}$ ($10^{17}$ cm)       & 4.83 & 4.83 & 4.83 & 4.83 & 4.83 \\
%$\tau_{\rm BLR}$                       & 0.1 & 0.1 & 0.1 & 0.1 & 0.1 \\
%$R_{\rm DT}$ ($10^{18}$ cm)            & 11.8 & 11.8 & 11.8 & 11.8 & 11.8 \\
%$\tau_{\rm DT}$                        & 0.1 & 0.1 & 0.1 & 0.1 & 0.1 \\
%$T_{\rm DT}$ (K)                       & 900 & 900 & 900 & 900 & 900 \\
$R_{\rm H}$ ($10^{17}$ cm)             & 1.6 & 1.6 & 1.6 & 1.6 & 1.6 \\
$B$ (G)                                & 6.5 & 7.2 & 5.8 & 7.3 & 7.8 \\
%$T_{\rm disk}$ (K)                     & 30000 & 30000 & 30000 & 30000 & 30000 \\
$P_{\rm rad}$ ($10^{44}$ erg s$^{-1}$)&2.99&5.06 &3.24&3.40&3.22\\
$P_e$ ($10^{44}$ erg s$^{-1}$)&0.90&1.75 &1.32&1.35&1.55\\
$P_p$ ($10^{46}$ erg s$^{-1}$)&1.74&3.08 &2.93&3.12&3.43\\
$P_B$ ($10^{46}$ erg s$^{-1}$)&0.89&1.20 &0.77&1.33&1.17\\
$P_{\rm jet}$ ($10^{46}$ erg s$^{-1}$)&2.66&4.36 &3.75&4.50&4.66\\
\bottomrule
\end{tabular*}
\end{table*}

%\begin{table*}[!htbp]
%\centering
%\caption{Energy densities and powers from SED modeling for the five activity periods (P1--P5).}
%\label{tab:jetset_energy_parameters}
%\small % Sets a uniform, readable font size
%\setlength{\tabcolsep}{0pt} % Let LaTeX handle spacing automatically
%\begin{tabular*}{\textwidth}{@{\extracolsep{\fill}}llccccc}
%\hline\hline
%Parameter&Description & P1 & P2 & P3 & P4 & P5 \\
%\midrule
%$U_e$ ($10^{-2}$ erg cm$^{-3}$)&Energy density of relativistic electrons&1.70 &2.98 &2.30 & 2.15 &3.19 \\
%$U_B$ (erg cm$^{-3}$)&Magnetic field energy density&1.68 &2.06 &1.33 &2.12 &2.42\\
%$P_{\rm rad}$ ($10^{44}$ erg s$^{-1}$)&Radiative jet power&2.99&5.06 &3.24&3.40&3.22\\
%$P_{\rm kin}$ ($10^{46}$ erg s$^{-1}$)&Kinetic power&1.75&3.10 &2.94&3.14&3.45\\
%$P_e$ ($10^{44}$ erg s$^{-1}$)&Power carried by electrons&0.90&1.75 &1.32&1.35&1.55\\
%$P_p$ ($10^{46}$ erg s$^{-1}$)&Power carried by protons&1.74&3.08 &2.93&3.12&3.43\\
%$P_B$ ($10^{46}$ erg s$^{-1}$)&Power carried by the magnetic field&0.89&1.20 &0.77&1.33&1.17\\
%$P_{\rm jet}$ ($10^{46}$ erg s$^{-1}$)&Total jet power&2.66&4.36 &3.75&4.50&4.66\\
%\bottomrule
%\end{tabular*}
%\end{table*}

\begin{figure*}[ht]
\centering

\begin{subfigure}{0.495\textwidth}
    \centering
    \includegraphics[width=\linewidth]{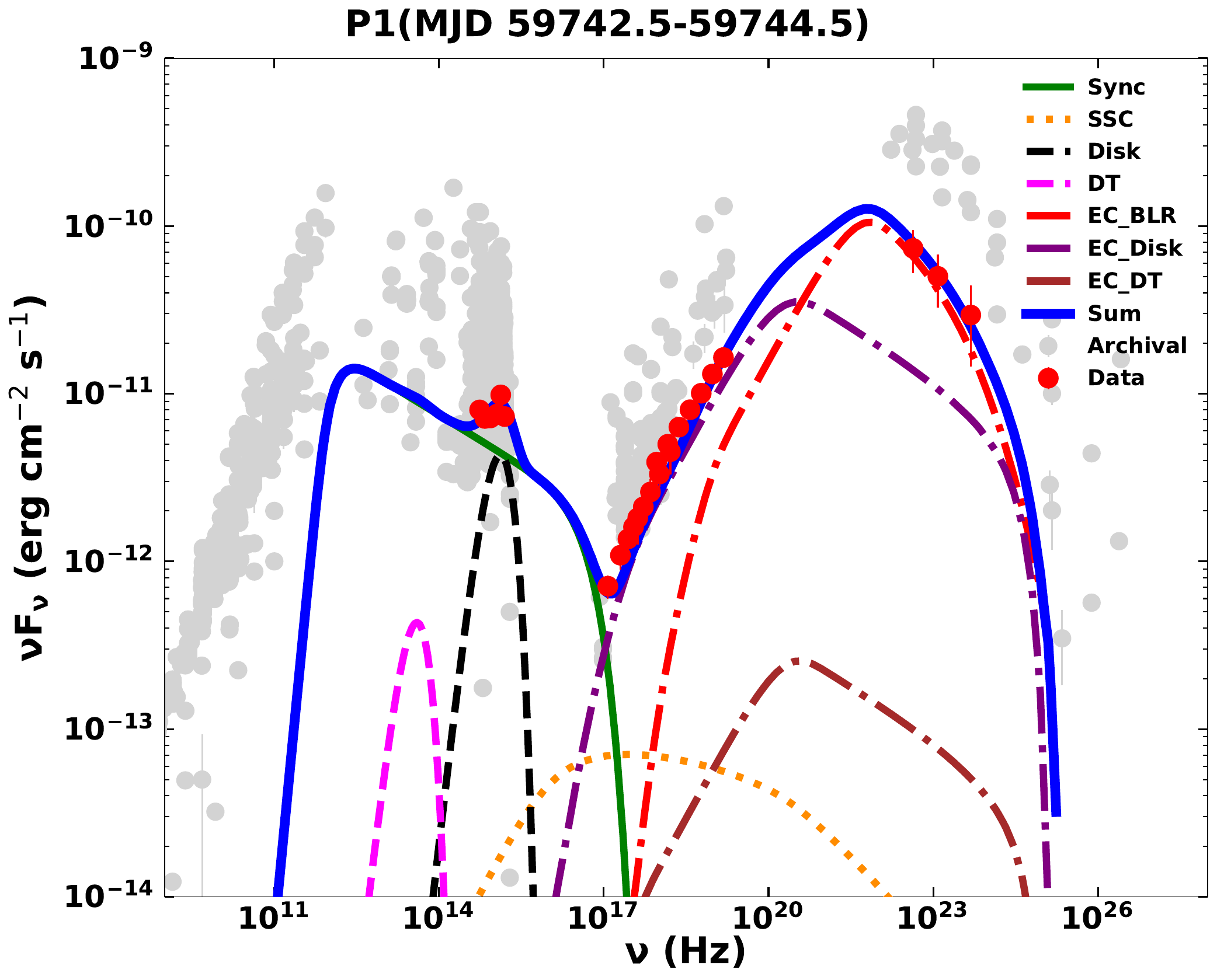}
\end{subfigure}
\hfill
\begin{subfigure}{0.495\textwidth}
    \centering
    \includegraphics[width=\linewidth]{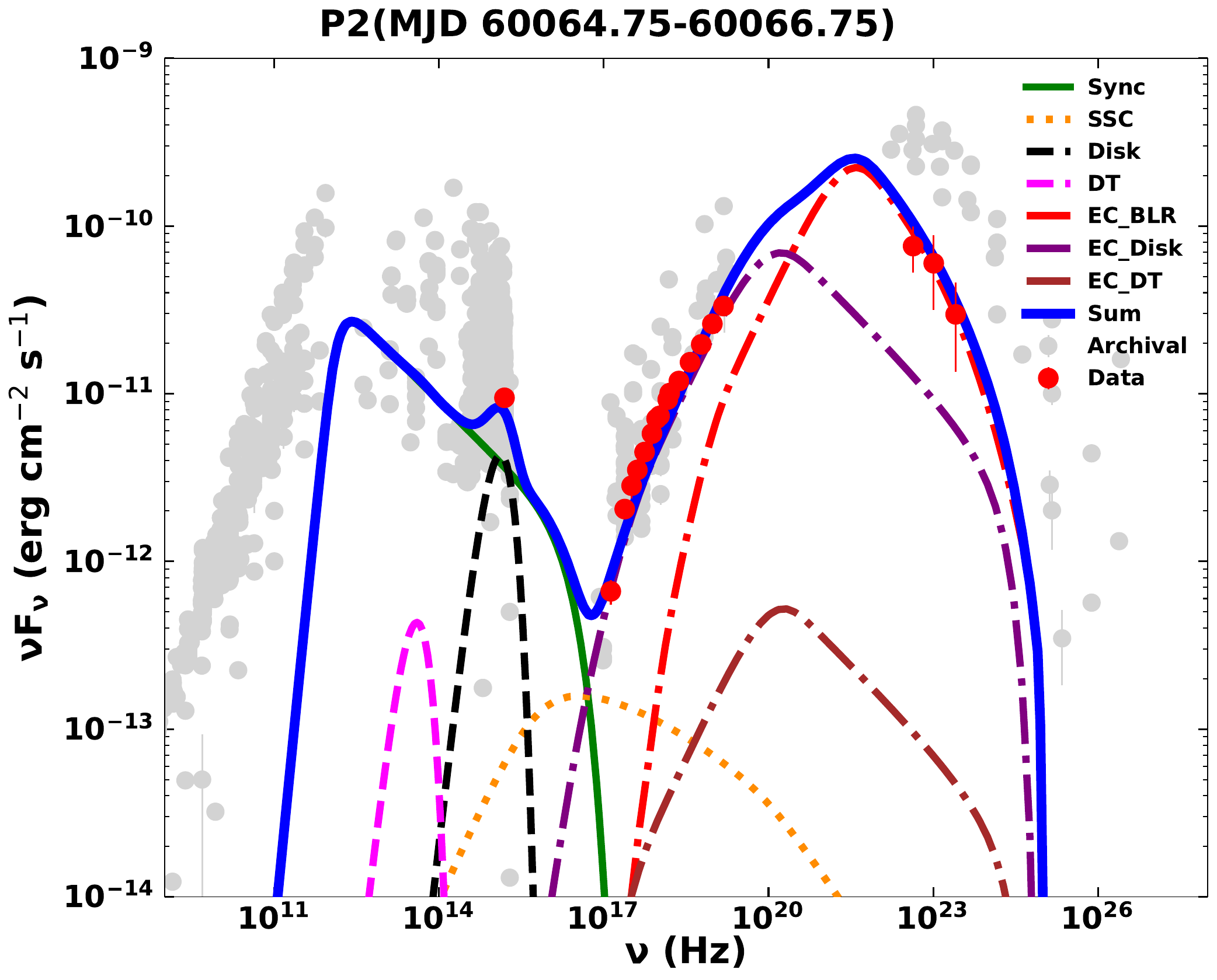}
\end{subfigure}

\vspace{0.2cm}

\begin{subfigure}{0.495\textwidth}
    \centering
    \includegraphics[width=\linewidth]{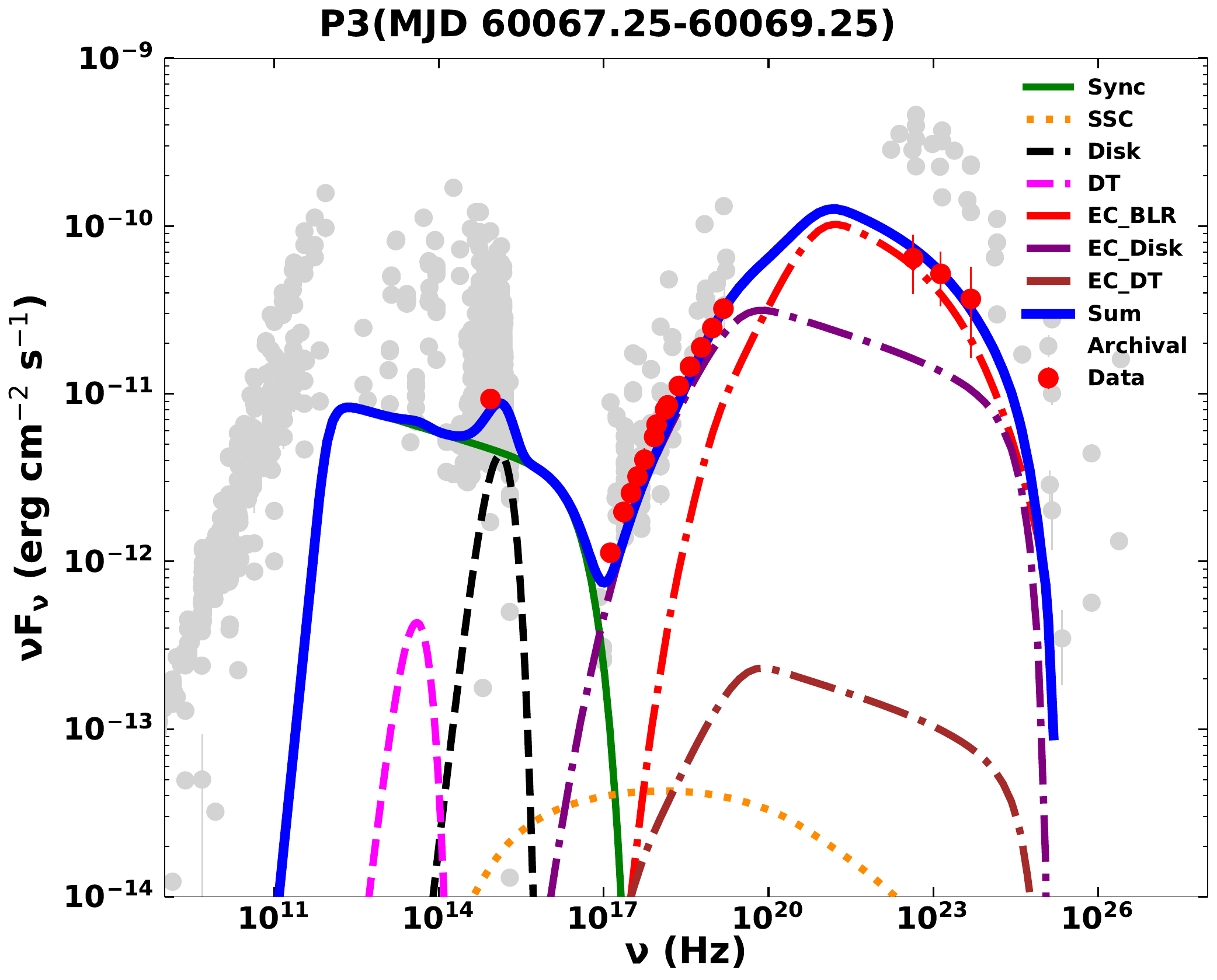}
\end{subfigure}
\hfill
\begin{subfigure}{0.495\textwidth}
    \centering
    \includegraphics[width=\linewidth]{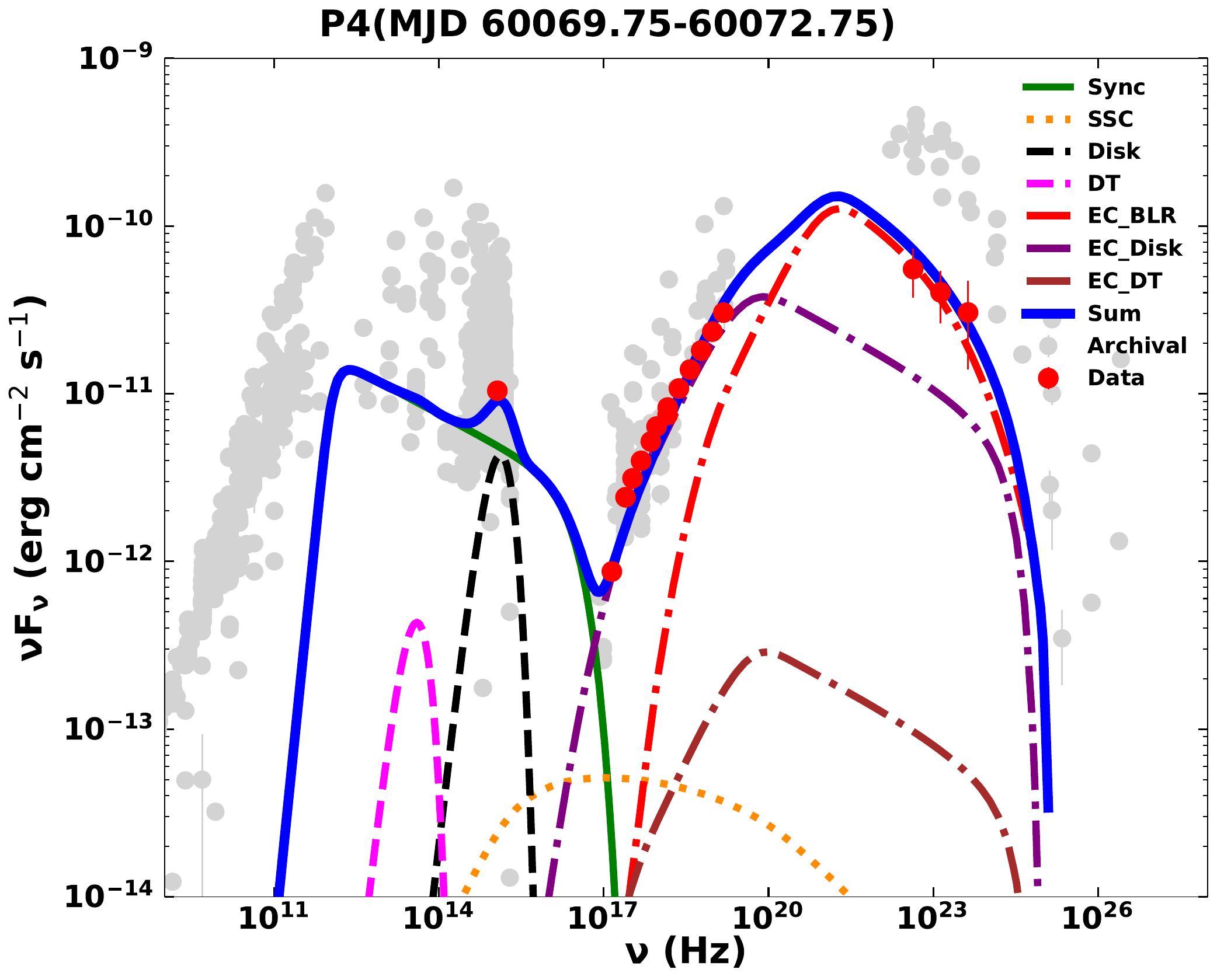}
\end{subfigure}

\vspace{0.2cm}

\begin{subfigure}{0.495\textwidth}
    \centering
    \includegraphics[width=\linewidth]{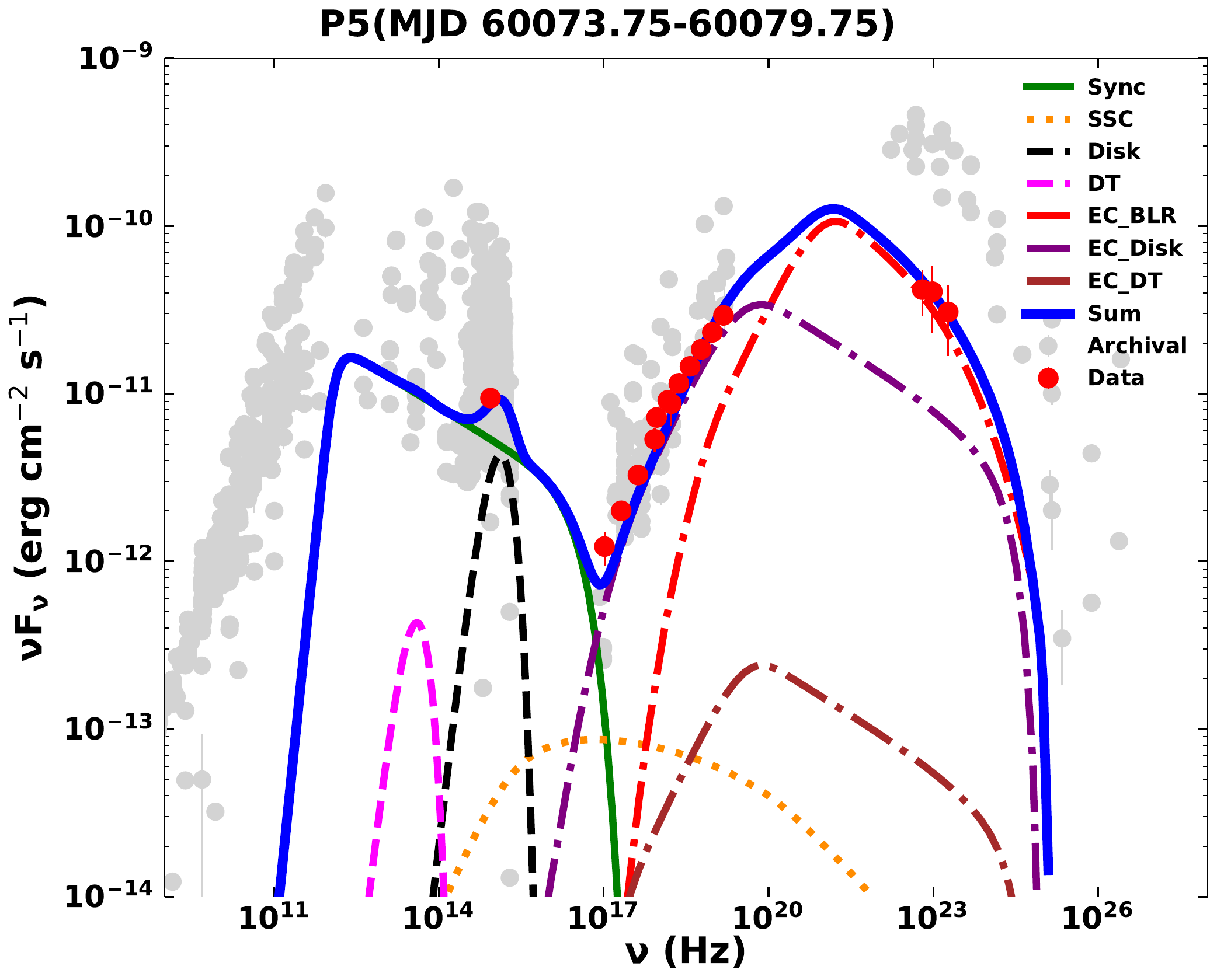}
\end{subfigure}

\caption{Plot showing the result of modeling the multi-wavelength SED for the five epochs. The thin green solid line shows the synchrotron, the dark orange dotted line shows SSC, the black dashed line refers to the blackbody emission from the disk, and the magenta dotted line shows the emission from the dusty torus. The red, purple, and brown dashed-dotted lines represent EC-BLR, EC-Disk, and EC-DT  models, respectively. The sum of all the radiative components is shown with the thick blue solid line.}
\label{fig:sed_5_epochs}
\end{figure*}

\section{Discussion}\label{sec:dis}
\subsection{Multi-wavelength Light Curves}
We analyzed two distinct observational windows for 3C 454.3: Period I and Period II, from the light curve. The maximum \(\gamma \)-ray flux densities recorded during these intervals were \(0.58 \times 10^{-6}\text{ ph cm}^{-2}\text{s}^{-1}\)
(Period I) and \(0.52 \times 10^{-6}\text{ ph cm}^{-2}\text{s}^{-1}\) (Period II). Because these peak values remain well below the usual flaring thresholds defined for this source (typically \(F > 1.0 \times 10^{-6}\text{ ph cm}^{-2}\text{s}^{-1}\)), both epochs are confirmed to represent a non-flaring, quiescent state of the blazar, aligning with long-term baseline studies (e.g., Sarkar et al. 2020). The light curve shows mild flux variation in different multi-band frequencies. The optical and UV bands show less variation compared to the X-ray and gamma-ray bands, especially during Period I. Therefore, we calculated the fractional variability amplitude \(F_{\text{var}}\) \citep{2003MNRAS.345.1271V} to study the flux variations across both time periods. Similar to typical blazars, the \(F_{\text{var}}\) increases with increasing frequency \citep{Soldi2008} during Period I. No significant variations were observed in the \(\gamma\)-ray and hard X-ray bands during Period II.

\subsection{Multi-wavelength SED}
We performed broadband SED modeling using the open-source Python framework \jetset. A one-zone leptonic \citep{1992ApJ...397L...5M} framework incorporating synchrotron and IC emissions was used to model the selected five epochs(P1--P5).  All the SEDs shown in Fig.~3 represent low-activity states, providing critical insights into the baseline physics of relativistic jets and accretion disks. Because 3C~454.3 is one of the most violently variable and luminous $\gamma$-ray sources in the sky, studying its calm phases allows astronomers to decouple steady-state thermal emission from overwhelming, non-thermal jet flares \citep{Raiteri_2008}. Analyzing this quiescent state is crucial for isolating the underlying steady-state energetics of the central engine. Furthermore, studying the quiescent phase is highly significant as it represents a system near equipartition and accurately reflects the long-term, steady-state behavior of the source. In the SED modeling, the ratio of particle-to-magnetic energy density ($U_e/U_b$) was found to drop below unity, indicating a slightly magnetically dominated environment \citep{2014Natur.515..376G,2016ApJ...828...13I}. This contrasts sharply with studies focused on strong flaring periods, where the jet becomes heavily particle-dominated due to rapid acceleration mechanisms like relativistic shocks or magnetic reconnection. Because these explosive, non-thermal processes are short-lived, the quiescent state provides a much better picture of the baseline jet magnetohydrodynamics. Moreover, by incorporating simultaneous \textit{NuSTAR} observations into our modeling, we achieve significantly tighter constraints on the high-energy spectrum, ensuring a robust, well-defined parameter space due to the broader multi-wavelength coverage. Typically, FSRQs exhibit minimal flux variation during their quiescent states \citep{2021MNRAS.504.1103R}, and similar steady trends are clearly visible in the underlying light curves presented in this work.

All SEDs exhibit a distinct peak in the optical-UV region, which arises from a significant contribution from the accretion disk along with the non-thermal synchrotron emission. Notably, the observed SED data points closely align with the accretion disk model, demonstrating that the spectroscopic data are highly consistent with the Big Blue Bump modeling \citep{Shang2005,Paliya_2021}. The optical polarization measurements taken during the \textit{IXPE} campaign revealed a negligible polarization \citep[][]{Marshall_2024}, which also supports the possible disk-dominated optical emission, as corroborated by the SED modeling done in this work.

In all the modeled SEDs, the X-ray emission is located on the low-energy side of the inverse Compton hump, indicating that the observed X-ray flux is associated primarily with inverse Compton (IC) processes rather than synchrotron radiation. The EC component provides the major contribution to the X-ray flux, while the SSC does not contribute significantly to the hard X-rays. The model reproduces the observed X-ray data reasonably well, indicating that inverse Compton scattering of ambient photons is an important mechanism for producing the X-ray emission. Furthermore, the EC component is expected to exhibit considerably lower X-ray polarization than pure synchrotron tracking \citep[][]{2013ApJ...774...18Z,2018ApJ...863...98P}, a physical prediction heavily supported by the recent non-detection of X-ray polarization observed by the \textit{IXPE} satellite \citep{Marshall_2024}. The high-energy ($\gamma$-ray) emission is consistently dominated by external Compton scattering of BLR photons (EC-BLR). The direct accretion disk photon field (EC-Disk) also contributes to the lower-energy side of the high-energy emission; however, as the frequency increases, its relative contribution decreases drastically. Contributions from the dusty torus (EC-DT) are comparatively weak. This hierarchy is a direct consequence of the localized geometric profile of the dissipation zone ($R_H \ll R_{\text{DT}}$); the relativistic Doppler-boosted seed photon field from the nearby surrounding BLR heavily outscales the dilute ambient photon density originating from the distant torus structure, naturally suppressing the secondary EC-DT spectral component. 

All detected $\gamma$-ray photons possessed energies below this 10~GeV threshold, and therefore they are capable of escaping the jet environment without suffering severe $\gamma$-$\gamma$ pair-production absorption against the dense local ambient radiation field \citep{2014ApJ...790..149L}. Consequently, our one-zone leptonic SED model predicts that the emission region is compact and resides within the interior cavity of the BLR.

\subsection{Comparison with Previous Analyses}

Most of our SED parameters are in excellent agreement with the quiescent state modeling of 3C 454.3 presented by \cite{2021MNRAS.504.1103R}. The primary discrepancies emerge in the disk luminosity (\(L_{\text{disk}}\)), the external radiation field boundaries (\(R_{\text{BLR}}\) and \(R_{\text{DT}}\)), and the low-energy electron spectral index (\(p\)). The differences in the Broad-Line Region distance and Dusty Torus radius are a direct mathematical consequence of lower disk luminosity used in our model, as these structural boundaries scale with \(\sqrt{L_{\text{disk}}}\). We observed a flat X-ray spectrum during the selected periods, which is consistent with the hard electron distribution at low energies, as reflected by the relatively low value of (\(p\)) obtained from our modeling.

When comparing our results with \cite{sahakyan2021modelling}, we find that both the UVOT peaks (\(6.3 \times 10^{14}\)~Hz) and the high-energy electron index (\(p_{1}\)) are in excellent agreement with their modeling. The magnetic field is also within the same range. However, the accretion disk luminosity in their model is \(4.57 \times 10^{46}\)~erg/s, which is almost double that of our value derived from optical spectroscopy. Our modeled broad-line region (BLR) properties do not align with theirs. Although the bulk Lorentz factors occupy a comparable range, the Doppler factors differ because our modeling explicitly incorporates a viewing angle constraint of \(1.3^{\circ }\). Regarding the particle distribution, our inferred values of \(p\) are close to 1.34, whereas their break Lorentz factor (\(\gamma_{b} = 240.7\)) is substantially larger than our inferred value. This difference may reflect different physical conditions and particle-acceleration properties in the respective emission regions. Finally, our inferred emission region size is slightly larger, which is consistent with the longer characteristic variability timescales associated with the non-flaring state.

When we compare our results  with \cite{2011ApJ...736L..38V} , our SED parameters (\(L_{\text{disk}}\), \(\Gamma \), and \(\theta \)) are nearly identical. Consequently, both models operate under the same external photon field baseline and experience equivalent relativistic Doppler boosting. However, our modeled emission region is significantly more compact, and the magnetic field strength is substantially larger than their findings; indicating different physical conditions within the emitting region. Furthermore, the Lorentz factors reported in their work (\(\gamma _{\text{min}}\), \(\gamma _{\text{max}}\), and \(\gamma _{\text{break}}\)) are significantly higher than our inferred values, while their electron spectral indices (\(p\) and \(p_{1}\)) are notably softer. Taken together, these microphysical discrepancies indicate that our resulting X-ray and \(\gamma \)-ray spectra are inherently harder than those presented in their work.

A comparison with \cite{bonnoli2011gamma} reveals that our primary parameters : the distance of dissipation zone (\(R_{\mathrm{H}}\)), the bulk Lorentz factor (\(\Gamma \)), and magnetic field strength-are in close agreement.  The high-energy electron spectral indices (\(p_1\)) are comparable between the two models, though the low-energy slope (\(p \)) is slightly harder than our values. Additionally, our maximum Lorentz factor (\(\gamma _{\max }\)) is larger and the break Lorentz factor (\(\gamma _{\text{break}}\)) is smaller compared to their findings. The combination of a lower \(\gamma _{\text{break}}\) and higher \(\gamma _{\max }\) indicates a different distribution of electrons across the high-energy portion of the particle spectrum.

A comparison of our broadband SED modeling results with \cite{2019MNRAS.486.1781R} reveals excellent agreement in the particle spectral indices (\(p\) and \(p_{1}\)), dusty torus temperature (\(T_{\mathrm{DT}}\)), and maximum electron Lorentz factor (\(\gamma _{\mathrm{max}}\)), suggesting broadly similar electron-distribution properties and external infrared photon-field conditions. However, our bulk Lorentz factor (\(\Gamma\)) and emission region radius (\(R\)) are larger than their reported values. Furthermore, our model yields a lower minimum Lorentz factor (\(\gamma _{\mathrm{min}}\)) and a significantly stronger magnetic field (\(B\)). These parameter differences suggest that, while the characteristic external radiation environment and upper limit of the electron energy distribution are broadly consistent between the two models, the physical conditions within the emitting region can differ. In particular, the stronger magnetic field in our model may lead to enhanced synchrotron cooling and consequently influence the shape of the particle and broadband emission spectra.

\section{Summary}\label{sec:summary}

We conducted a detailed multiwavelength analysis of the extraordinary blazar 3C~454.3 during Period I (MJD 59740-59747) and Period II ( MJD 60064-
60080). The broadband SED was modeled using the \texttt{JetSet} package. The key results of our analysis are summarized below:
\begin{enumerate}
    %\item The hard X-ray emissions (3--79~keV) observed by \textit{NuSTAR} during these epochs represent the first-ever \textit{NuSTAR} detection of 3C 454.3.
    \item During the \textit{IXPE} campaign, the blazar exhibited only mild flux variations across the electromagnetic spectrum.
    %\item The multi-wavelength light curve shows minimal flux variations in the low-frequency regime but exhibits significant variability at higher frequencies.
    \item The low-energy component of the SED is well-described by synchrotron radiation, with an additional optical ``big blue bump'' accounting for the thermal luminosity of the accretion disk. In the high-energy regime, the X-ray spectrum is successfully modeled by the external Compton (EC) scattering of accretion disk photons, which is supported by the non-detection of X-ray polarization observed by the \textit{IXPE} satellite. The $\gamma$-ray spectrum is modeled via the external Compton scattering of BLR photons.
    \item Because all detected $\gamma$-ray photons possessed energies below 10~GeV, our one-zone leptonic SED model predicts that the emission region is compact and resides well within the BLR.
    \item A comparison of our best-fit \jetset SED modeling parameters with similar studies provides valuable insights into the underlying particle acceleration mechanisms and jet dynamics in FSRQs.
\end{enumerate}

\section*{Acknowledgments}

We gratefully acknowledge the use of data from the \textit{Fermi} Science Support Center (FSSC), \textit{Swift} data, and \textit{NuSTAR} data from the High Energy Astrophysics Science Archive Research Center (HEASARC) at NASA's Goddard Space Flight Center, as well as archival data from the Space Science Data Center of the Italian Space Agency (ASI). Sajeesh George sincerely appreciates the research facilities provided by UGC-SAP and FIST-II (SR/FIST/PSI-159/2010) (DST, Government of India) in the Department of Physics, University of Calicut. We also extend our gratitude to the Department of Physics, Govt. College Kodanchery, Calicut.

\section*{Data Availability}
The data utilized in this study are publicly accessible and were obtained from the archives available at \url{https://Fermi.gsfc.nasa.gov/}.

\appendix
\par\vspace{0.5cm}
\bibliographystyle{elsarticle-harv} 
%\bibliographystyle{ieeetr}
%\bibliographystyle{harv} 
%\bibliographystyle{elsarticle-num-names} 
% Loading bibliography database
\bibliography{VHE}
\end{document}